\documentclass[letter]{aa} 

\usepackage{graphicx}
\usepackage{txfonts}
\begin{document}

   \title{The shape evolution of cometary nuclei via anisotropic mass loss}

   \subtitle{}

   \author{D. E. Vavilov,
          \inst{1}
          S. Eggl,
          \inst{2}
          Yu. D. Medvedev
          \inst{1}
          	\and
          P. B. Zatitskiy
          \inst{3,4}          
          }

   \institute{Institute of Applied Astronomy of the Russian Academy of Sciences,
              St. Petersburg, Russia\\
              \email{vavilov@iaaras.ru}
            \and
 				Department of Astronomy, University of Washington, 
 				Seattle, WA, USA \\
 				\email{eggl@uw.edu}
 			\and			
 				Chebyshev Laboratory, St. Petersburg State University, 
 				St. Petersburg, Russia             
         	\and
 				St. Petersburg Department of V.A. Steklov Institute of Mathematics of the Russian Academy of Sciences, St. Petersburg, Russia 
             }

   \date{Received ; accepted }

 
  \abstract
   {Breathtaking imagery recorded during the European Space Agency's Rosetta mission confirmed the bilobate nature of comet 67P/Churyumov-Gerasimenko's nucleus. Its peculiar appearance is not unique among comets. The majority of cometary cores imaged at high resolution exhibit a similar build. Various theories have been brought forward as to how cometary nuclei attain such peculiar shapes.}
   {We illustrate that anisotropic mass loss and local collapse of subsurface structures caused by non-uniform exposure of the nucleus to solar irradiation can transform initially spherical comet cores into bilobed ones.}
   {A mathematical framework to describe the changes in morphology resulting from non-uniform insolation during a nucleus' spin-orbit evolution is derived. The resulting partial differential equations that govern the change in the shape of a nucleus subject to mass loss and consequent collapse of depleted subsurface structures are solved analytically for simple insolation configurations and numerically for more realistic scenarios.}
   {The here proposed mechanism is capable of explaining why a large fraction of periodic comets appear to have peanut-shaped cores and why light-curve amplitudes of comet nuclei are on average larger than those of typical main belt asteroids of the same size.}
   {}


   \maketitle
%

\section{Introduction}

\begin{figure}
	\centering
	\includegraphics[width=1\linewidth]{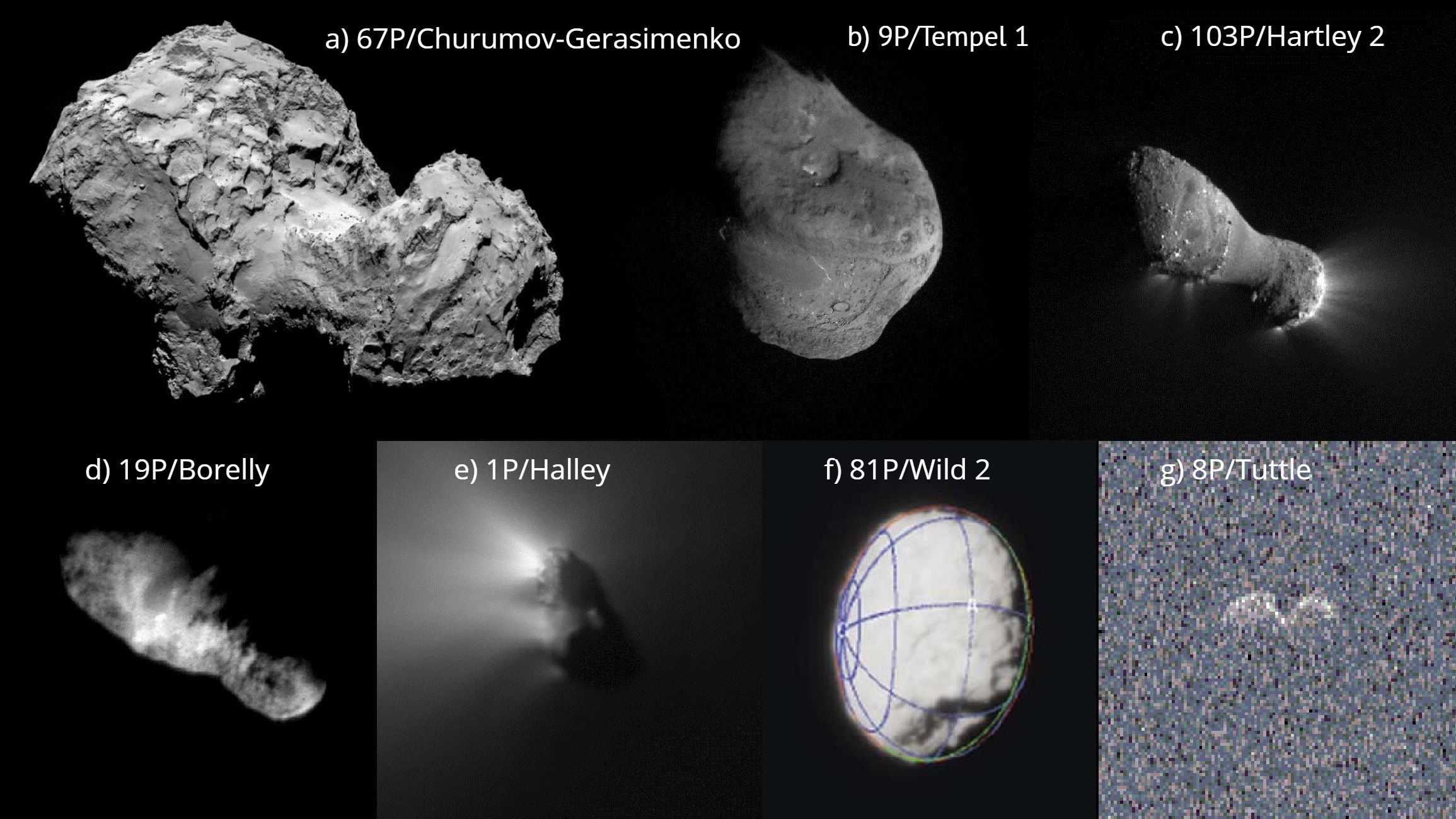}
	\caption{Cometary nuclei imaged from spacecraft encounters or ground-based radar. a) 67P/Churyumov–Gerasimenko. Image credit: ESA/Rosetta/MPS for OSIRIS Team MPS/UPD/LAM/IAA/SSO/INTA/UPM/DASP/IDA. b) 9P/Tempel 1. Image credit: PIA02142, Courtesy NASA/JPL-Caltech. c) 103P/Hartley 2. Image credit: PIA13570, Courtesy NASA/JPL-Caltech. d) 19P/Borelly. Image credit: PIA03500, Courtesy NASA/JPL-Caltech. e) 1P/Halley. Image credit: ESA/MPS. f) 81P/Wild 2, see~\citep{2004JGRE..10912S02D}. g) 8P/Tuttle. Image credit: Arecibo Observatory Planetary Radar, resolution $1 \  \mathrm{\mu s} \times 0.5$~Hz, see~\citep{2010Icar..207..499H}.}
	\label{Fig:Comets}%
\end{figure}

Bilobed configurations appear to be common in cometary nuclei. The detailed imagery recorded in the framework of the Rosetta mission saw 67P/Churyumov-Gerasimenko join the ranks of dumbbell-shaped comets alongside 103P/Hartley 2, 19P/Borrelly, 1P/Halley and 8P/Tuttle~\citep{2015A&A...583A..34K,2015Natur.526..402M,2016A&A...592A..63D}, as shown in Fig.~\ref{Fig:Comets}. Current hypotheses as to how such peculiar shapes emerge range from high speed collisions of proto-comets~\citep{2017A&A...597A..61J, 2017A&A...597A..62J, 2017MNRAS.465.3949B} over fission and reconfiguration~\citep{2018NatAs...2..379S, 2016Natur.534..352H} to a slow merging of two separately formed proto-cores on similar orbits~\citep{2015Sci...348.1355J}. Apart from collisions, gradual mass loss through sublimation has long been considered a possible mechanism to shape cometary nuclei~\citep{2003AJ....125.3366J}. The surface of 67P, for instance, shows significant outgassing as well as changes to its surface when exposed to solar radiation~\citep{2015Sci...347a1044S}. The corresponding loss of water-ice mixed with other chemical compounds can be substantial for short period comets that approach the Sun on a regular basis. The surface thickness of 67P, for instance, is believed to shrink on average by $1.0 \pm 0.5 \ \mathrm{m}$ per orbit~\citep{2015A&A...583A..38B}.  Exposure to sunlight can vary over a comet's surface depending on its geometry and spin state~\citep{2015Sci...347a1044S}. How small differences in local insolation, sublimation and outgassing rates sculpt cometary nuclei over their lifetime is one of the key concerns of this work. 

Comets are believed to form in the outer Solar System as kilometer sized single-lobed objects~\citep{2016A&A...592A..63D}. On the journey towards the inner Solar System the outer shell of a cometary nucleus is dehydrated and accumulates coatings rich in organic compounds on top of a layer of ice~\citep{2016Natur.529..368F, 2016Sci...354.1566F}. The nucleus is also heated periodically by incident sunlight subject to the core's rotation. Due to the periodic heating the nucleus' subsurface layers experience depletion of cryomagma and consequent structural collapse leading to a localized compactification of core material~\citep{2016Icar..272..356M}.  {Even if a newly formed cometary core was largely homogenous to begin with, the afore mentioned processes would foster radial disparity in the nucleus' core. The denser, outer shell depleted of volatiles is less likely to allow cometary material to be lost at elevated rates. The material closer to the core retains a high volatile content and a relatively low bulk density.} 

Should the nucleus lose some of its shell the local differences in mass loss rates could change the shape of the comet's core in a non-trivial fashion. Subsurface material could be exposed, for instance, as a consequence of non-catastrophic collisions~\citep{2018NatAs...2..379S}. Collision probabilities for comets in the outer solar system are hard to quantify, however. In this work, we, thus, focus on the more general concept of anisotropic mass loss.


\section{Carving cometary nuclei through anisotropic mass loss}

\begin{figure}
   \centering
   \includegraphics[width=1\linewidth]{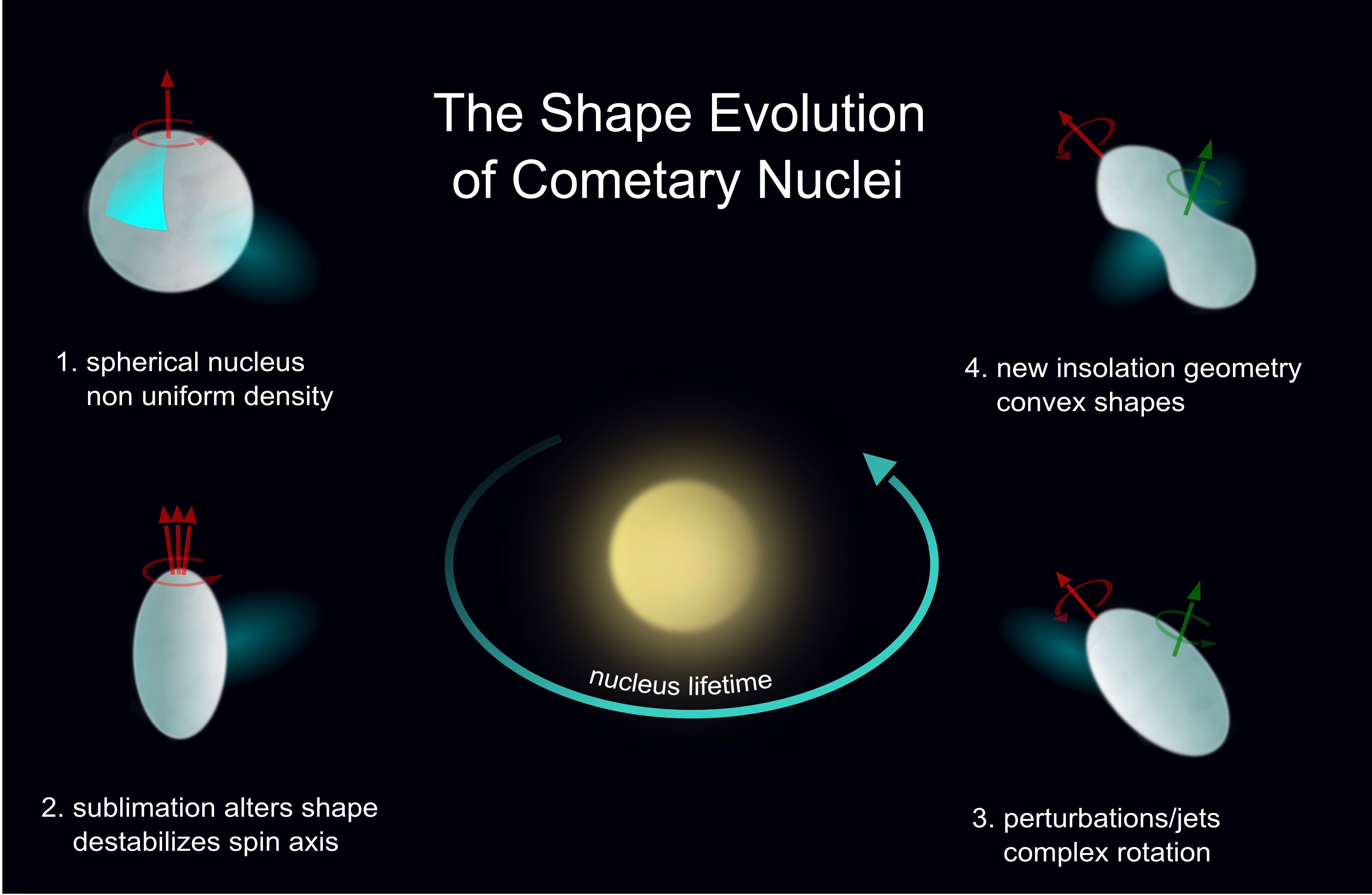}
   \caption{Shape and spin evolution of cometary nuclei can lead to bilobed shapes.}
              \label{Fig:ShapeEvolution}%
    \end{figure}

For comets in principle axis rotation the energy input from the sun is highest around the nucleus' subsolar manifold, the path on the nucleus' surface that experiences the largest insolation over one spin period. Since the  {penetration depth of the sublimation front} into the core is proportional to the local energy input, mass loss rates in those regions are enhanced compared to the rest of the core. Nuclei with spin axis orientations perpendicular to their orbital plane, for instance, are bound to lose matter more quickly around their ''waist'' (Fig.~\ref{Fig:ShapeEvolution}). The collapse of desiccated subsurface structures and formation of sinkholes can bring the new surface layer closer to more pristine material in the center of the core~\citep{2016Icar..272..356M}. The now volatile rich subsurface once again yields higher mass loss rates. Over time, this process of  {anisotropic mass loss} can reshape a comet's nucleus into a more elongated form.   

The more elongated affected cometary cores become, however, the more the principle moment of inertia decreases around the nucleus' primary axis of rotation. Torques induced by localized outgassing such as jets or close encounters with planets can then destabilize the spin axis leading to the complex rotation states observed in the majority of comets. Mass loss continues to alter the shape of the comet even in complex rotation states. Our simulations show that if a comet's distance to the sun remains large enough to avoid disintegration within a few orbits, a wide variety of geometries of cometary nuclei arises naturally (Fig.~\ref{Fig:ModelComets}). How quickly  {anisotropic mass loss} changes the shape of a comet depends on the comet's size and spin-orbit evolution (see Appendix~\ref{App:SecTime}). 

Forming bilobate shapes such as 67P's through  {the proposed mechanism} requires the rate of mass loss in the outer shell of the cometary core to be smaller than that that closer to the interior of the primordial body.  {Otherwise, if the mass loss rate is homogeneous or decreases towards the core's center the comet's shape keeps being convex.} Given the processing and desiccation the outer shell experiences over the lifetime of a comet, in particular before the latter enters the inner solar system, one would expect precisely such a configuration~\citep{2016Icar..272..356M}.  Mass loss rates of material closer to the core remain high due to the higher volatile content. If 67P formed as a single, roughly spherical object, the so-called ''neck'' (Hapi) region would have been close to the primordial core and Hapi does show enhanced outgassing activity compared to other domains on 67P~\citep{2015Sci...347a1044S}.  We would like to stress that the higher outgassing rates near Hapi were not caused by enhanced insolation at the time of observation. In fact, the neck received around 15\% less sunlight than other regions~\citep{2015Sci...347a1044S}. Higher mass loss rates around the neck, thus, point towards compositional or structural differences compared to the rest of the comet, which is in line with our model.

 {One of the main arguments against 67P being an evolved state of a once monolithic body is based on the identification of distinct geological landmarks on both lobes identified as non-matching ''strata''~\citep{2015Natur.526..402M}. Once  {anisotropic mass loss} has carved out sufficient material to form a neck, however, processes such as rotational fracturing and recombination are more likely to occur, and they may lead to a reconfiguration of the lobes affecting current strata orientation~\citep{2016Natur.534..352H}. Consequently, we argue that mismatches in strata orientation do not exclude monolithic formation scenarios.}

\section{Modeling anisotropic mass loss based shape changes}
\subsection{Mathematical formulation}

In order to understand how the anisotropic loss of material affects the shape of a comet's nucleus over time we first derive a simplified mathematical model. The aim of that model is to qualitatively describe the changes in a nucleus' morphology resulting from non-uniform insolation during a nucleus' spin-orbit evolution. The following simplifying assumptions allow us to derive the differential equations that govern the change in the shape of a nucleus subject to mass loss and consequent collapse of depleted subsurface structures:

\begin{enumerate}
	\item A comet's nucleus forms as a roughly spherical body.  
	\item The nucleus spins rapidly enough, so that we can average anisotropic mass loss induced shape changes over one spin period.
	\item The changes in the comet's shape occurring over one orbital period are small~\citep{2017Sci...355.1392E}.
	\item 
     {Due to the desiccated shell and volatile rich core the nucleus' acquired through its journed towards the inner solar system, the mass loss rate is assumed to be non-uniform and increases towards the center of the nucleus.}
\end{enumerate}

We represent the comet's shape by a closed surface in spherical coordinates $S(\phi, \psi)$. The partial differential equation for the nucleus' surface then reads:

\begin{equation}
\frac{\partial S(\phi, \psi,t)}{\partial t} = -\frac{\bar{Z}}{\cos \theta} \label{eq:1},
\end{equation}

\noindent where $\theta$ is the angle between the surface normal at point $(\phi, \psi)$ and the cometocentric radial vector to point $(\phi, \psi)$ and $\bar{Z}$ is the mass loss rate averaged over, both orbital and spin period. The derivation of the above equation is presented in Appendix~\ref{App:SecPartdiffeq}. For simple configurations equation~(\ref{eq:1}) permits analytical solutions that are discussed in Appendix~\ref{App:SecAnalSol}). More realistic cases require numerical treatment. To find numerical solutions we applied an algorithm that passed time reversibility and convergence tests when compared to analytical solutions for simple geometries.

\subsection{Non-constant mass loss throughout the nucleus}
Results from the Rosetta Radio Science Investigation experiment~\citep{2016Natur.530...63P} suggest that 67P has a largely homogenous density distribution.  Nevertheless, higher mass loss rates originating from the Hapi region, the so-called ''neck'' of 67P have been observed~\citep{2015Natur.526..402M, 2015Sci...347a1044S}. We identify the neck as a more pristine core of a roughly spherical, primordial nucleus and, based on the afore mentioned observations, assume a general dependency of mass loss rates on the cometocentric distance, $R$.  {Here we assume that the center of the comet's nucleus contains primordial matter that sublimates about twice as fast as the processed matter near the surface.} More precisely, the mass loss rate through a unit area perpendicular to a surface element shall be given by a  {relative} function  $\Gamma(R)$,  {which is normalized by the loss rate at the center of the comet}. In order to simulate the behavior of a compact, desiccated shell, we used the following relations:

\begin{displaymath}
\Gamma(R) = \left[
\begin{array}{rl}
1, & R \leq 0.5 \\
-5R+3.5, & R \in [0.5,0.6] \\
0.5, & R \geq 0.6
\end{array} \right . ,
\end{displaymath}

\noindent where $R$ is the cometocentric distance in units of the comet's initial radius.  {Here we have been conservative in assuming that processing of cometary material happens down to approximately half of the core's radius. This means more time is required to change the shape of the comet significantly, since the thicker the shell the lower the average (bulk) mass loss rate.}
As a consequence, the mass loss rate increases towards the center of the comet. We found, however, that our results are robust against changes in the precise form of $\Gamma(R)$, as long as the mass loss rate is lower in the outer shell of the comet's core.

\subsection{Influence of complex rotation}

Apart from the heliocentric distance and local mass loss rates, the perhaps most influential parameter that shapes a comet's core is its spin. We assume the original nucleus' spin axis is tilted with respect to the orbital angular momentum axis by the angle $\alpha_1$, the comet's initial obliquity. 
After the comet's shape and with it its principle moments of inertia have been sufficiently altered by anisotropic mass loss, perturbations such as the sudden onset of jets, torques due to the sun, close approaches with planets, or collisions with interplanetary debris would lead to an eventual destabilization of the primordial rotation state.
We model this effect and the resulting complex rotation of the comet by adding a second spin axis perpendicular to the first half way through the simulation. The latter mimics the change in the principle moments of inertia leading to a more stable rotation with respect to the new short axis. The angle between the second spin axis and the orbital angular momentum vector is named $\alpha_2$.

\section{Results}

\begin{figure}
	\centering
	\includegraphics[width=1\linewidth]{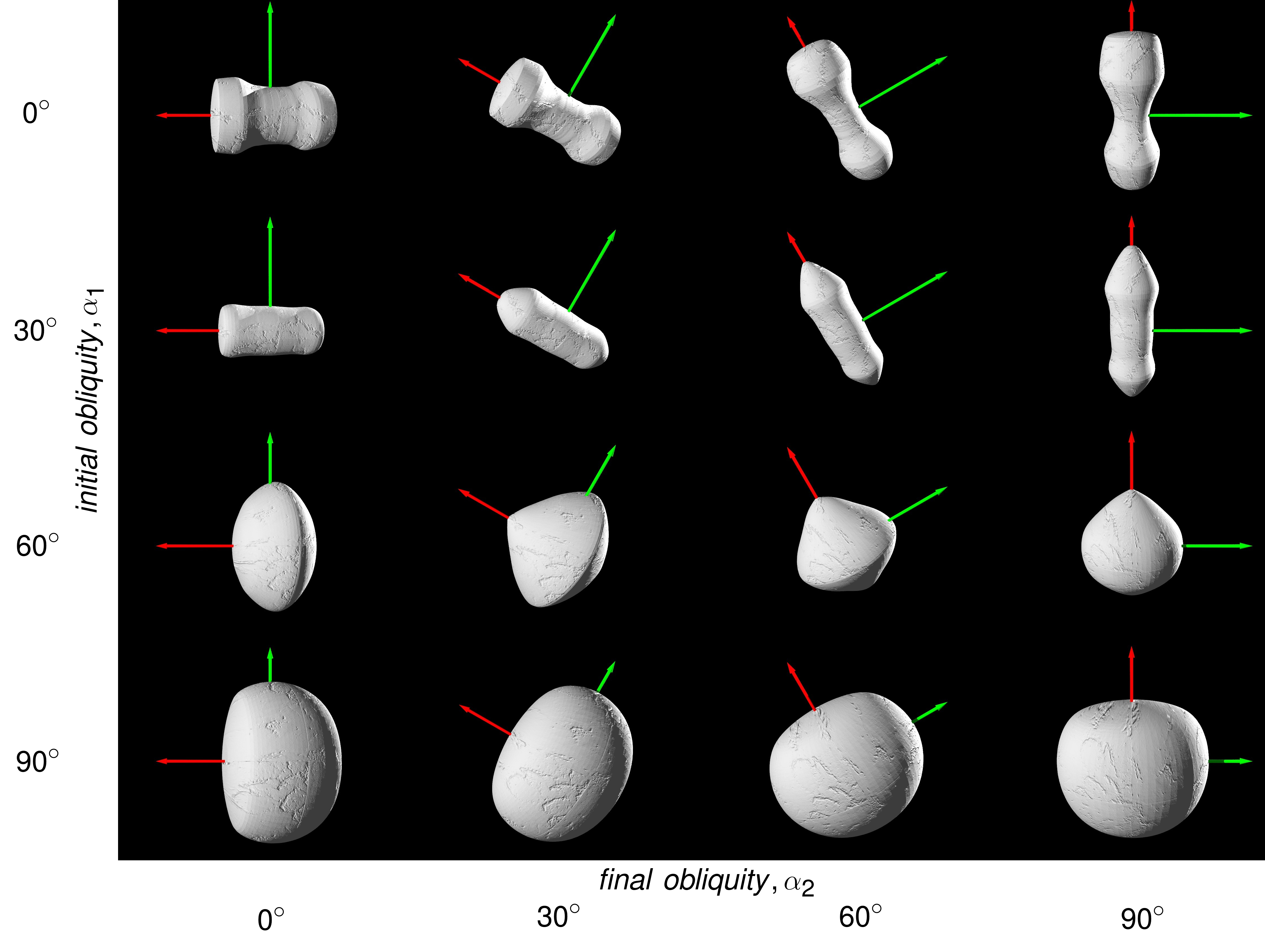}
	\caption{Possible shapes of cometary nuclei as a result of anisotropic mass loss are shown for a variety of configurations. The outcome is determined by the nucleus' initial obliquity before substantial mass loss sets in ($\alpha_1$) and final obliquity after destabilization of the original spin state ($\alpha_2$).  The corresponding spin axes are colored in red and green, respectively. The nuclei in this graph are oriented according to their final obliquity with respect to the orbital angular momentum (vertical direction).}
	\label{Fig:ModelComets}%
\end{figure}

If the shape of a comet's nucleus can be understood as a result of its particular spin-orbit history it is of interest to see whether some shapes are more likely to occur than others. 
Fig.~\ref{Fig:ModelComets} in the main text shows the outcome of our simulations for a set of angles $\alpha_1$ and  $\alpha_2$, where $\alpha_1 \in \{0^{\circ}, 30^{\circ},60^{\circ},90^{\circ}\}$ and $\alpha_2 \in \{0^{\circ}, 30^{\circ},60^{\circ},90^{\circ}\}$ for a comet experiencing orbitally averaged insolation values that correspond to a circular orbit at 3 au. 

Sampling a broad range of spin-orbit configurations we find that of all the investigated spin histories of cometary cores more than 70\% produced elongated shapes. Cometary nuclei were found to become bilobate in roughly half of all cases. This result is in agreement with the sample of imaged cometary nuclei taking into account small number statistics, where four out of six comets visited by spacecraft have bilobate characteristics~\citep{2018AJ....155..246N, 2011Sci...332.1396A}. Comparing Fig.~\ref{Fig:Comets} and Fig.~\ref{Fig:ModelComets} shows that practically all of the observed morphologies of cometary nuclei are mirrored in the simulation. Even the ''pan-cake'' shape of 81P can emerge naturally as a consequence of  {anisotropic mass loss}.

That elongated shapes are the most common outcome of mass loss induced evolution processes is confirmed by observational data. Comets tend to have larger light curve amplitudes than asteroids --- a telltale sign of the more elongated shapes of the former~\citep{1988ApJ...328..974J, 2004come.book..223L}. 
The here presented model, thus, not only correctly predicts the wide variety of shapes encountered in cometary nuclei, it also explains observed differences in the light curve amplitudes of asteroids and cometary nuclei. 

How long would it take for a comet with a radius of $1 \ \mathrm{km}$ on an orbit similar to 67P to be split into two parts by anisotropic mass loss? Inserting the corresponding quantities into equation~(\ref{App:Eq:Time}) we find that  $T_l=14250$~years (see Appendix~\ref{App:SecTime}). Over one orbital period such a comet is bound to lose mass equivalent to a $0.5 \ \mathrm{m}$ thick shell. The former value is consistent with the averaged per orbit shrinkage of 67P calculated from Rosetta spacecraft observations, namely $1.0 \pm 0.5 \ \mathrm{m}$, see~\citep{2015A&A...583A..38B}. In contrast, several million years of anisotropic mass loss would be necessary for main belt comets to experience a similar change in shape.

\section{Summary}

   \begin{enumerate}
      \item Anisotropic mass loss caused by non-uniform exposure to sunlight can carve cometary cores on timescales comparable to residence times in the inner solar system. 
      \item While the processes responsible for reshaping a nucleus are complex and merit a more detailed investigation, our simplified model is capable of explaining why the majority of comets exhibits elongated, often bilobate shapes.   
      \item The fact that  {anisotropic mass loss} would be more effective in shaping comets rather than asteroids which tend to have lower volatile contents may also explain why observations of cometary nuclei yield comparatatively larger light-curve amplitudes. 
      \item  {Mismatching "strata" observed on the lobes of 67P may be a consequence of the formation of a "neck" at the center of the comet through anisotropic mass loss followed by rotational fracturing of the comet and eventual recombination and reconfiguration of the lobes.}  
   \end{enumerate}

\begin{acknowledgements}
      The here presented research was made possible through funding from the Russian Scientific Foundation, project number 16-12-00071.
\end{acknowledgements}

%
%

\bibliographystyle{aa}
\bibliography{references}

\begin{appendix} 
	\section{Partial differential equation derivation}
	\label{App:SecPartdiffeq}
\begin{figure}
	\centering
	\includegraphics[width=1\linewidth]{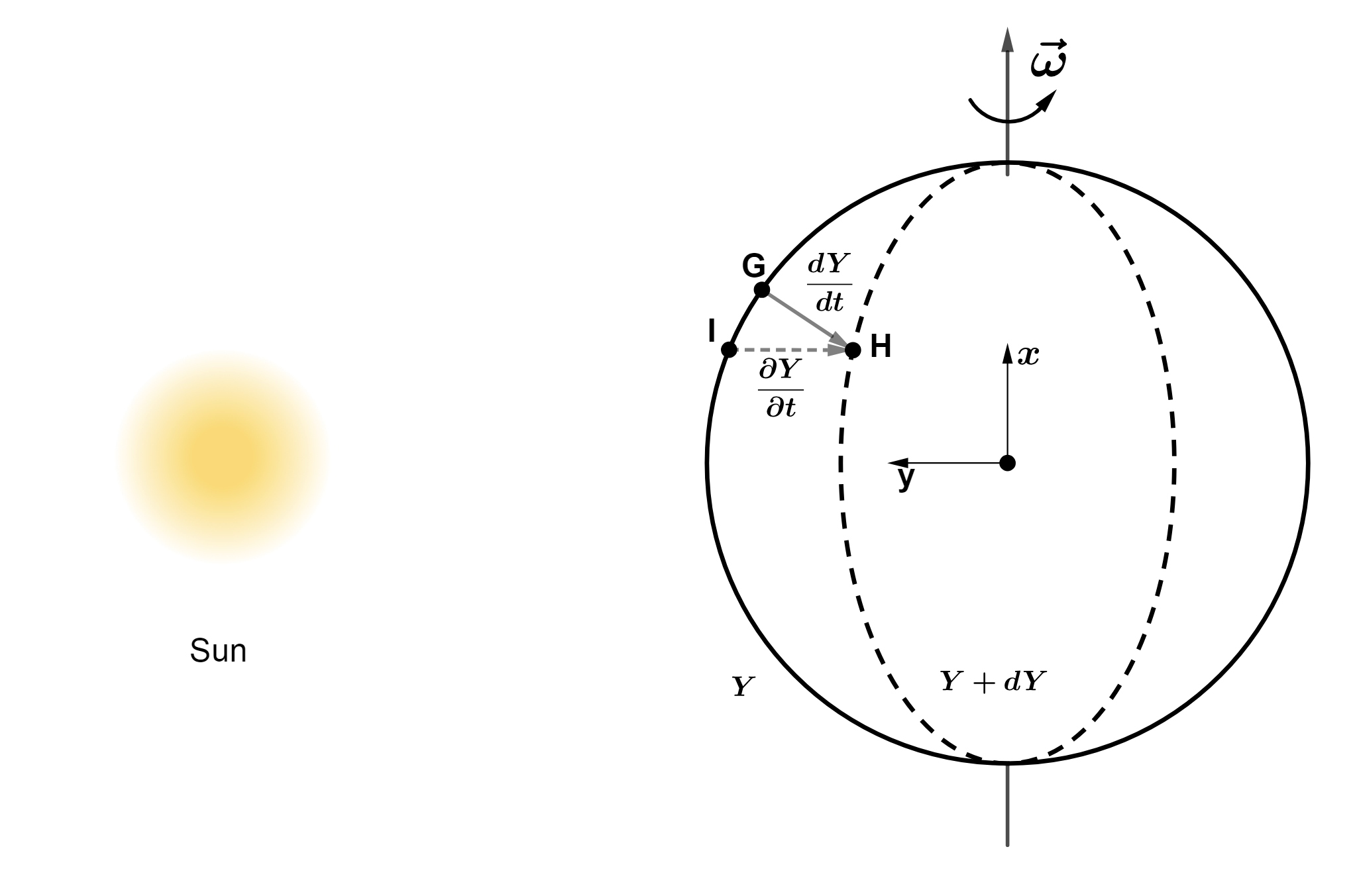}
	\caption{Anisotropic mass loss induced shape changes as described in equation~(\ref{App:eq:part}). I and G are points on the comet's original surface while H lies on the processed surface resulting from volatile loss, desiccation and collapse of subsurface cavities.}
	\label{App:Fig:AML}%
\end{figure}

Let us assume that the spin axis of the nucleus is perpendicular to the orbital plane. In that case the core's shape will be axially symmetric. An infinitesimally thin slice of the comet's nucleus containing the spin axis can then be modeled via a function Y(x) that traces the silhouette of the nucleus. The y-axis of the corresponding coordinate system is oriented towards the sun and the x-axis coincides with the spin axis (see Fig.~\ref{App:Fig:AML}).

Changes in $Y(x)$, thus, correspond to changes in the nucleus' shape. A collapse of local subsurface structures would shift a small area along the normal to $Y(x)$ towards the center of the comet. Let $\theta$ be the angle between the direction of insolation and the normal to the curve $Y(x)$ and let $Z$ be the mass loss rate. If insolation values on surface elements change with time $Y(x)$ becomes $Y(X(t),t)$  and the change in $Y$ due to volatile loss and consequent collapse of subsurface elements is given by 
\begin{displaymath}
\frac{dY}{dt} = -Z \cos \theta; \ \frac{\partial X}{\partial t} = - Z \sin \theta.
\end{displaymath}
\noindent With $\frac{dY}{dt}=\frac{\partial Y}{\partial t}+\frac{\partial Y}{\partial X}  \frac{\partial X}{\partial t}$   taking into account $\frac{\partial Y}{\partial X}= -\tan \theta$  we find:
\begin{equation}
\frac{\partial Y}{\partial t} = -\frac{Z}{\cos \theta}
\label{App:eq:part}
\end{equation}
$Z$ is a function of $\cos \theta$ and the heliocentric distance of the nucleus, $r$.  According to assumption number 4, $Z$ also depends on the cometocentric distance, $R$. Since $\cos \theta=(1+(\frac{\partial Y}{\partial X})^2 )^{-1/2}$, equation~(\ref{App:eq:part}) is a partial differential equation. How the mass loss rate  $Z$ depends on $\cos \theta$ and $r$ depends on the so-called ''sublimation function''. The latter parameterizes how much matter sublimates from a unit area perpendicular to the incident sunlight over unit time as a function of the comet's heliocentric distance. In this work we use a sublimation function for water ice derived by Sekanina~\citep{1992acm..proc..545S, 1973AJ.....78..211M}
\begin{displaymath}
g(r)=0.111262 \left(\frac{2.808}{r}\right)^{2.15} \left(1+\left[\frac{r}{2.808}\right]^{5.093} \right)^{-4.6142},
\end{displaymath}
\noindent where $r$ is heliocentric distance in astronomical units. In order to account for the fact that not every surface element of the cometary nucleus is necessarily perpendicular to the incoming sunlight the mass loss rate has to be proportional to $g(r_0/\sqrt{\cos \theta})$, where $r_0$ is the heliocentric distance of a unit sized surface element. The mass loss rate averaged over spin period ($\hat{Z}$) then equals:
\begin{displaymath}
\hat{Z}(\theta, R, r_0) = \Gamma(R)\frac{1}{2\pi}\int^{\pi/2}_{-\pi/2} g(r_0/\sqrt{\cos \theta \cos \alpha}) \ d\alpha
\end{displaymath}
\noindent where  $\Gamma(R)$ describes the relative mass loss rate in the nucleus as a function of the core radius $R$ and $\alpha$ is the azimuthal angle between the radius vector pointing towards the surface element and the direction of insolation. For nuclei with non-constant density,  $\Gamma(R)$ would be inversely proportional to the density distribution. If the changes in the comet's shape are relatively minor over one orbit, $\hat{Z}(\theta, R, r_0)$ can finally be averaged over one orbital period to obtain the absolute mass loss rate $\bar{Z}(\theta, R)$
\begin{displaymath}
\bar{Z}(\theta, R) = \int^{2\pi}_{0} \hat{Z}(\theta, R, r_0) \ dM,
\end{displaymath}
\noindent where $M$ is the mean anomaly of the comet's orbit. 

\subsection{Analytical solution. Constant mass loss rates throughout the nucleus}
\label{App:SecAnalSol}
For constant mass loss rates throughout the core $\Gamma(R) = 1$, equation~(\ref{App:eq:part})  can be solved analytically. With $\bar{Z}(\theta, R) = \Phi \left( \frac{\partial Y}{\partial X} \right)$ the solution reads:
\begin{displaymath}
\left\{ \begin{array}{ll}
X(u,t) = & t\Phi'(u) - \frac{u}{\sqrt{1+u^2}} \\
Y(u,t) = & ut\Phi'(u) - t\Phi(u) + \frac{1}{\sqrt{1+u^2}} \\
\end{array}
\right . .
\end{displaymath}
For non-constant mass loss rates in the nucleus ($\Gamma(R) \neq 1$) solutions to the shape evolution equations have to be found numerically. In that case all derivatives are approximated by finite differences that can be solved with a suitable algorithm. 

\subsection{Equation for arbitrary rotation }
If the spin axis is not perpendicular to the orbital plane we represent the comet's shape by a closed surface in spherical coordinates $S(\phi, \psi)$. The partial differential equation for the nucleus' surface then reads:
\begin{displaymath}
\frac{\partial S(\phi, \psi,t)}{\partial t} = -\frac{\bar{Z}}{\cos \theta},
\end{displaymath}
\noindent where $\theta$ is the angle between the surface normal at point $(\phi, \psi)$ and the cometocentric radial vector to point $(\phi, \psi)$. 
The results obtained by the fully numerical solution are in good agreement with the analytical solution for those cases where the spin axis is perpendicular to the orbital plane. Forward and consequent backward propagation in time returns the initial shape with negligible errors.

\section{Mass loss timescale}
\label{App:SecTime}
In order to determine how long it would take cometary nuclei to evolve into the observed shapes through anisotropic mass loss we estimate the mass-loss near the equator of a comet with zero obliquity. The mass loss flux per unit area reads 
\begin{displaymath}
Z(r) = Z_0 g(r),
\end{displaymath}
where $r$ is the heliocentric distance and $Z_0$ represents the evaporation flux at a heliocentric distance of 1 au where $Z_0 = 3 \cdot 10^{17} \mathrm{molecules \ cm^{-2} sec^{-1}}$~\citep{1973AJ.....78..211M}. The mass loss flux averaged over one rotation period of the nucleus is:
\begin{displaymath}
\hat{Z}(r) = \frac{Z_0}{2\pi} \int_{-\pi}^{\pi} g(r/\sqrt{\cos \phi}) \ d\phi.
\end{displaymath}
If the rotation period of the comet is much shorter than the comet's orbital period and spin orbit resonances can be excluded, we can average the mass loss flux over one orbital period as well so that
\begin{displaymath}
\bar{Z}(a,e) = \int_{0}^{2\pi} \hat{Z}(r(M)) \ dM,
\end{displaymath}
\noindent where $a$ and $e$ are the semi-major axis and the eccentricity of the comet's orbit and $M$ is the mean anomaly. The number of molecules in any given column of cometary material is 
\begin{displaymath}
N = p \frac{N_A m}{w},
\end{displaymath}
\noindent $N_A$  being the Avogadro constant, $m$ is the mass of the column. Since the overwhelming part of the sublimated material is water ice, $w$ shall denote the molar mass of water, and $p$ is a coefficient that models the nucleus' mean porosity. A rough estimate for the mass loss timescale of a comet on a heliocentric orbit with semimajor axis $a$ and eccentricity $e$ is 
\begin{equation}
T_l = \frac{N}{\bar{Z}(a,e)}.
\label{App:Eq:Time}
\end{equation}

\end{appendix}

\end{document}